\documentclass[letterpaper]{article} % DO NOT CHANGE THIS
\usepackage{xxxx2026}  % DO NOT CHANGE THIS
\usepackage{times}  % DO NOT CHANGE THIS
\usepackage{helvet}  % DO NOT CHANGE THIS
\usepackage{courier}  % DO NOT CHANGE THIS
\usepackage[hyphens]{url}  % DO NOT CHANGE THIS
\usepackage{graphicx} % DO NOT CHANGE THIS
\usepackage{natbib}  % DO NOT CHANGE THIS AND DO NOT ADD ANY OPTIONS TO IT
\usepackage{caption} % DO NOT CHANGE THIS AND DO NOT ADD ANY OPTIONS TO IT
\usepackage{amsmath}
\usepackage{amssymb}
\usepackage{mathtools}
\usepackage{amsthm}
\theoremstyle{plain}

\theoremstyle{definition}

\theoremstyle{remark}

\usepackage{multirow}
\usepackage{arydshln}
\usepackage{colortbl}
\definecolor{Gray}{gray}{0.95}

\usepackage{booktabs}

\usepackage{amsmath}
\usepackage{amssymb}
\usepackage{mathtools}
\usepackage{amsthm}

\theoremstyle{plain}
\theoremstyle{definition}
\theoremstyle{remark}

\usepackage{multirow}
\usepackage{arydshln}
\usepackage{colortbl}
\definecolor{Gray}{gray}{0.95}
\usepackage{booktabs}

\title{RAGAR: Retrieval Augmented Personalized Image Generation \\ Guided by Recommendation}
\author{
    Run Ling \textsuperscript{\rm 1,\rm 7} \thanks{These authors contributed equally to this work.},
    Wenji Wang\textsuperscript{\rm 1} \footnotemark[1],
    Yuting Liu\textsuperscript{\rm 1},
    Guibing Guo\textsuperscript{\rm 1}\thanks{Corresponding author: alice@example.com},
    Haowei Liu\textsuperscript{\rm 2},
    Jian Lu\textsuperscript{\rm 2},
    Quanwei Zhang\textsuperscript{\rm 3},
    Yexing Xu\textsuperscript{\rm 4},
    Shuo Lu\textsuperscript{\rm 5},
    Yun Wang\textsuperscript{\rm 6},
    Yihua Shao\textsuperscript{\rm 5},
    Zhanjie Zhang\textsuperscript{\rm 3},
    Ao Ma\textsuperscript{\rm 7},
    Linying Jiang\textsuperscript{\rm 1},
    Xingwei Wang\textsuperscript{\rm 1}
}
\affiliations{
    \textsuperscript{\rm 1} Northeastern University, China \\
    \textsuperscript{\rm 2} Chongqing University of Post and Telecommunications\\
    \textsuperscript{\rm 3} Zhejiang University\\
    \textsuperscript{\rm 4} Sun Yat-sen University \\
    \textsuperscript{\rm 5} Institute of Automation, Chinese Academy of Sciences \\
    \textsuperscript{\rm 6} City University of Hong Kong \\
    \textsuperscript{\rm 7} JD.com\\
    runling001@outlook.com,
    \{wenjiwang, liuyuting\}@stumail.neu.edu.cn, 
    guogb@swc.neu.edu.cn
}

\usepackage{bibentry}
\begin{document}

\maketitle

\begin{abstract}
    Personalized image generation is crucial for improving the user experience, as it renders reference images into preferred ones according to user visual preferences. Although effective, existing methods face two main issues. First, existing methods treat all items in the user's historical sequence equally when extracting user preferences, overlooking the varying semantic similarities between historical items and the reference item. Disproportionately high weights for low-similarity items distort user visual preferences for the reference item. Second, existing methods heavily rely on consistency between generated and reference images to optimize generation, which leads to underfitting user preferences and hinders personalization. To address these issues, we propose \underline{R}etrieval \underline{A}ugmented Personalized Image \underline{G}ener\underline{A}tion guided by \underline{R}ecommendation (RAGAR). Our approach uses a retrieval mechanism to assign different weights to historical items according to their similarities to the reference item, thereby extracting more refined users' visual preferences for the reference item. Then we introduce a novel rank task based on the multi-modal ranking model to optimize the personalization of the generated images instead of forcing depend on consistency. Extensive experiments and human evaluations on three real-world datasets demonstrate that RAGAR achieves significant improvements in both personalization and semantic metrics compared to five baselines.
\end{abstract}

\section{Introduction} \label{sec:intro}
Personalized image generation has been widely applied in scenarios such as advertising systems and chat software. It aims to render reference images into preferred ones based on the user's visual preferences.
Existing image generation methods \cite{glide,ldm,ti} often produce similar outputs for different users with similar input, failing to meet users' needs for personalization.
As a personalized generation method, PMG~\cite{pmg} extracts user preferences from user historical sequences by treating each item equally, ignoring the varying semantic similarities between the reference item and historical items, such as semantic differences in categories. This disproportionately amplifies the influence of irrelevant items, causing the extracted user preference to deviate from the true visual preferences on the reference item.
Fig. \ref{fig:toy example}(a) shows a comparative example, if a user historical sequence includes electronic products and clothes, and the reference item is "short sleeves", an overly simplistic approach might generate a striped T-shirt with the Apple logo----an outcome that misses the user's true preferences. By incorporating semantic-based retrieval, more relevant visual preferences, such as "clothing, sport, pink", can be captured. Hence, we propose the assumption that filtering the items semantically related to the reference item enhances user preference modeling.

\begin{figure}[t]
    \centering
    \includegraphics[width=\linewidth,scale=1.00]{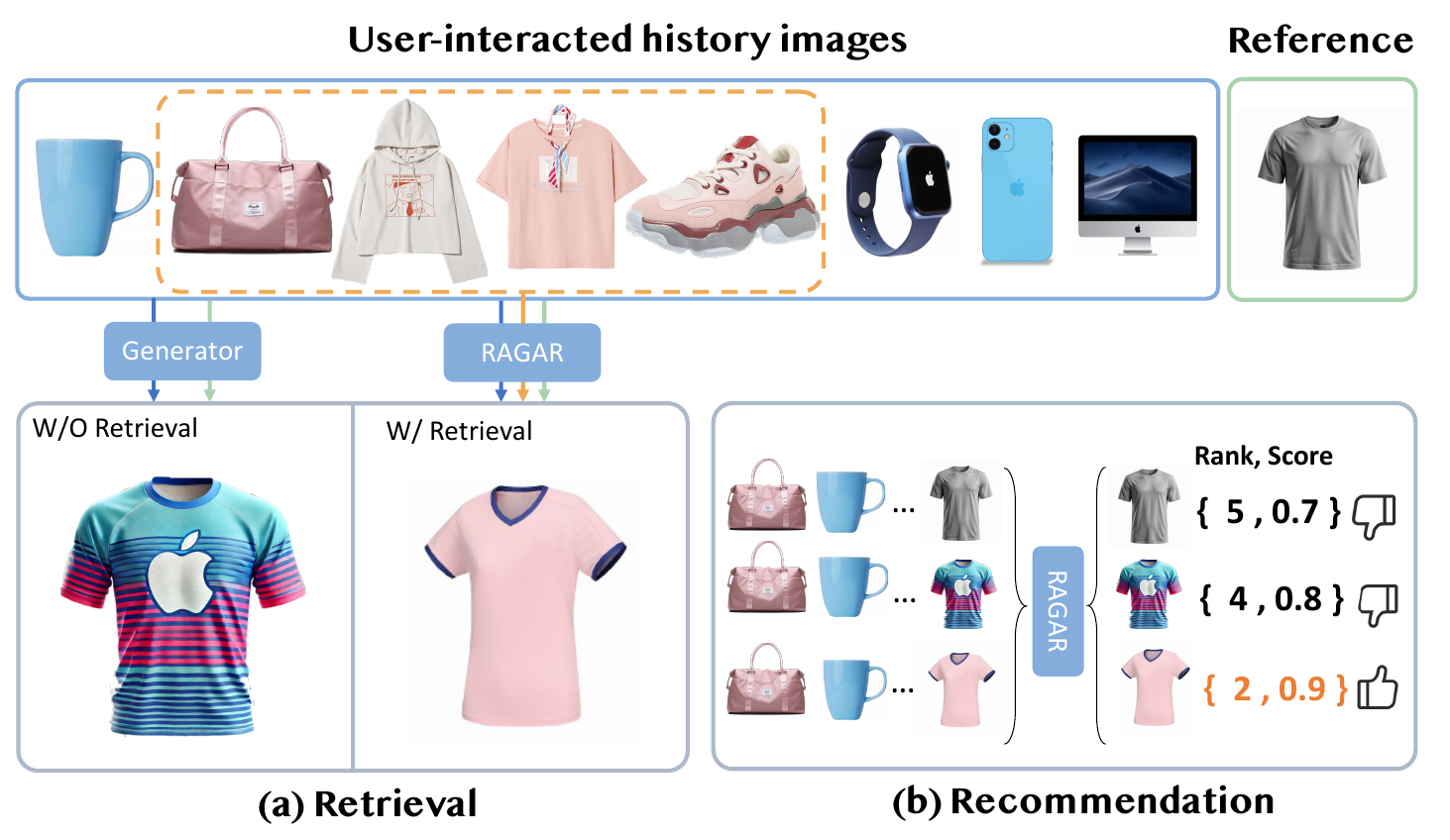}
    \caption{\label{fig:toy example}\textbf{Effect of retrieval and recommendation on image quality.} Lower rank and higher score of RAGAR-generated images suggest a strong positive correlation between retrieval/recommendation and personalized generation quality.}
    \label{fig:toy example} 
\end{figure}

\begin{figure}[h]
    \centering
    \includegraphics[width=\linewidth,scale=1.00]{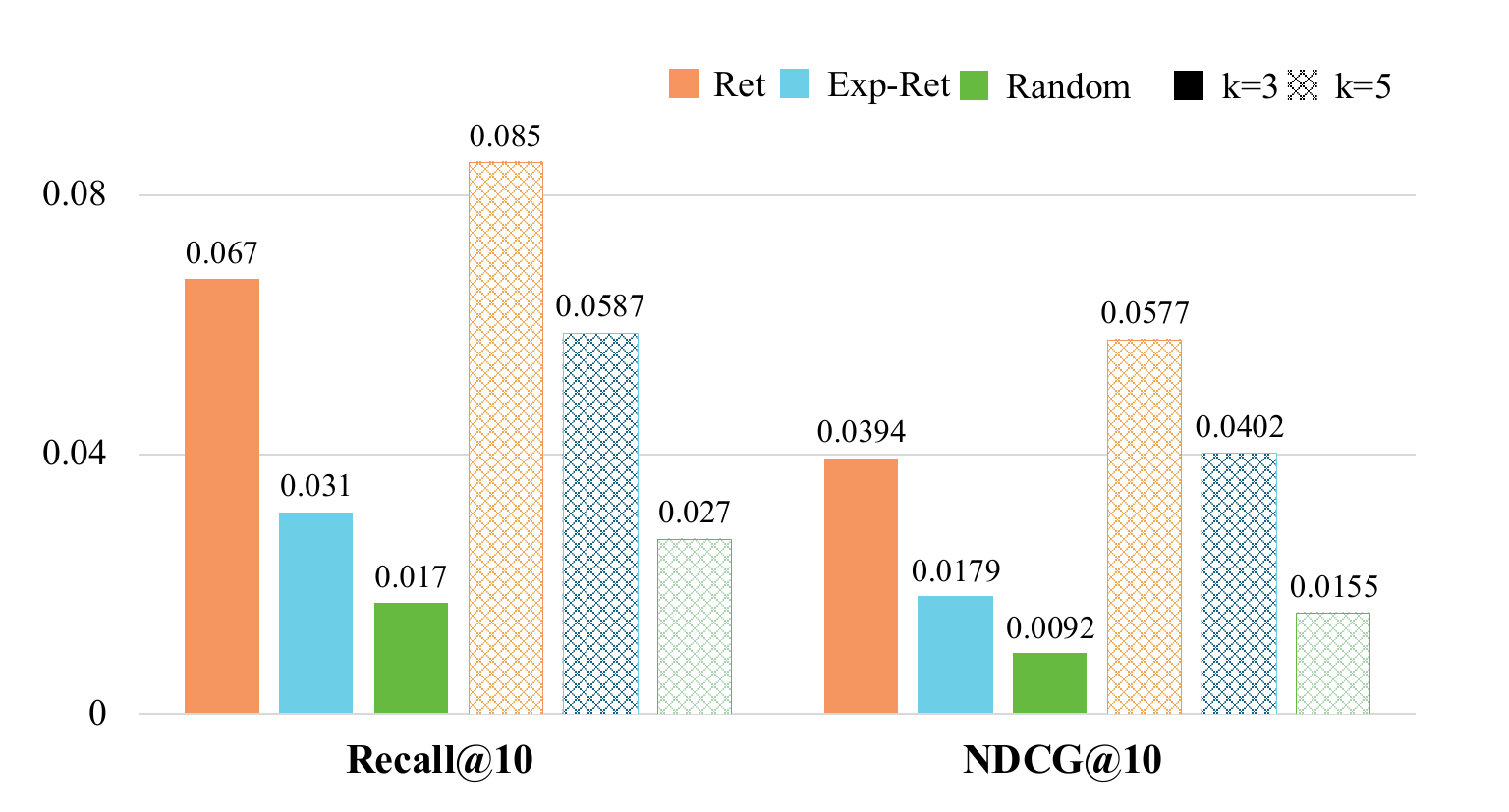}
    \caption{\label{fig:valide_res}\textbf{Comparison of user history sequence construction strategies on POG.} Top-$k$ retrieved items (Ret) consistently outperform lower-ranked (Exp-Ret) and random (Random) ones, validating that high-ranking relevant items better capture user preferences.}
    \label{fig:valide_res}
\end{figure}

% \begin{figure}[t]
% \centering
% \begin{minipage}[t]{0.48\linewidth}
%     \centering
%     \includegraphics[width=\textwidth]{images/toy_example_sm.pdf}
%     \caption{\label{fig:toy example}\textbf{Effect of retrieval and recommendation on image quality.} Lower rank and higher score of RAGAR-generated images suggest a strong positive correlation between retrieval/recommendation and personalized generation quality.}
%     \label{fig:toy example} 
% \end{minipage}
% \hfill
% \begin{minipage}[t]{0.48\linewidth}
%     \centering
%     \includegraphics[width=\textwidth]{images/valid_res_single.pdf}
%     \caption{\label{fig:valide_res}\textbf{Comparison of user history sequence construction strategies on POG.} Top-$k$ retrieved items (Ret) consistently outperform lower-ranked (Exp-Ret) and random (Random) ones, validating that high-ranking relevant items better capture user preferences.}
%     \label{fig:valide_res}
% \end{minipage}
% \vspace{-1.7em}
% \end{figure}

Another challenge lies in defining and evaluating personalization quantitatively.  Current methods rely on either human evaluation or large multi-modal models \cite{dsg,gpt4v4t3d}, both of which are resource-intensive. Recent work~\cite{pmg} evaluates and rectifies generated images heavily based on their pixel similarity to the reference image, resulting in limited personalization. Inspired by advances in recommendation systems, which excel in learning and assessing personalized preferences, we propose using multi-modal recommendation models to evaluate images, shown in Fig. \ref{fig:toy example}(b). These models provide an efficient and effective way to measure personalization. Moreover, by combining models and metrics from the recommendation domain, we propose and validate two assumptions: 1) items selected through retrieval are more closely aligned with user preferences than randomly selected items; 2) retrieving items semantically related to the reference item enhances the learning of user preferences compared to irrelevant items. We compare three types of user historical sequences on POG~\cite{POG} dataset: \textbf{Ret} contains the top-$k$ items selected with the retrieval method described in Sec. \ref{sec:retrieval module}. \textbf{Exp-Ret} (Expanded Retrieval) contains items ranked between $(k, 2k]$ by the retrieval method. \textbf{Random} contains $k$ randomly selected items. The results are shown in Fig. \ref{fig:valide_res}. The Random sequence has the lowest values in metrics, demonstrating that randomly selected items are less effective for preference modeling. Our first assumption holds because both Ret and Exp-Ret achieve higher metrics, suggesting that items related to the reference items are crucial for modeling preferences. Moreover, both metrics of Ret are higher than those of Exp-Ret, which implies the second assumption holds. Therefore, both of our assumptions hold in practical applications, motivating us to retrieve higher-ranked items to enhance the preferences modeling. More results can be found in the Appendix.
% ~\ref{Assumpation Validation}. 

% Hence, the paper proposed the Retrieval Augmented Personalized Image Generation guided by Recommendation (i.e., \textbf{RAGAR}), comprising three key modules: The \textbf{Retrieval module} retrievals historical items based on semantic by correlation unit and fuses visual features by fusion unit to form a retrieval-augmented preference. The \textbf{Generation module} summaries item keywords from the sequence and integrates detailed and global preferences from balance calibrator to generate images. The \textbf{Relection module} employs multi-assessments to guide the model to balance personalization and semantic fidelity.

Hence, we propose \textbf{RAGAR} (Retrieval Augmented Personalized Image Generation guided by Recommendation), comprising three key modules. The \textbf{Retrieval module} identifies semantically relevant historical items via a correlation unit and aggregates their visual features through a fusion unit to construct a retrieval-augmented preference. The \textbf{Generation module} extracts item keywords from the user sequence and integrates both detailed and global preferences, calibrated by the balance calibrator, to generate personalized images. Finally, the \textbf{Reflection module} applies multiple assessment signals to guide the model toward achieving a better balance between personalization and semantic fidelity.

The main contribution of the paper can be summarized as follows:
\begin{itemize}
\setlength{\itemsep}{0.5em}    % 控制item之间的间距
\setlength{\parskip}{0.3em}    % 控制段落之间的间距
\item To our knowledge, we are the first to emphasise and explore the intrinsic relationship between historical and reference items. We hypothesise this relationship is crucial in capturing preferences more accurately and guiding the personalized generation process. To support this hypothesis, we conduct a comprehensive data analysis on real-world datasets, which reveals significant impact of item-to-item relevance on personalization quality.
\item We propose a novel personalized image generation framework, RAGAR, which is specifically designed to address the dual objectives of semantic alignment and personalization. We utilize the retrieval assumption realized by calculating the semantic similarity between items to enhance semantic consistency. Additionally, we incorporate a discriminator to assess and enhance image quality, promoting better personalization.
\item Experiments on three diverse and real-world datasets demonstrate that RAGAR outperforms five strong baseline methods across multiple evaluation metrics. The empirical results highlight the effectiveness of our retrieval-based design and the discriminator.
\end{itemize}

\section{Methods}
\subsection{Task Formulation}

We formulate the task of personalized image generation in a multi-modal setting. Let $u$ denote a user, and let their historical interaction sequence be represented as: $\mathbf{S}_u = \{I_1, I_2, \dots, I_{N-1}\}$, where each item $I_j = \{\text{img}_j, \text{txt}_j\}$ is a multi-modal pair consisting of an image $\text{img}_j$ and a corresponding textual description $\text{txt}_j$. The final item $I_N = {\text{img}_N, \text{txt}_N}$ serves as the reference item, which specifies the semantic content for generation. The goal is to generate a personalized image $\hat{\text{img}}_u$ that reflects the user's historical preferences captured in $\mathbf{S}_u$ and aligns semantically with the reference item $I_N$.
Formally, the task is to learn a generation function:
\begin{equation}
\hat{\text{img}}_u = \mathcal{G}(\mathbf{S}_u, I_N)
\end{equation}
such that $\hat{\text{img}}_u$ maximizes personalization fidelity with respect to $\mathbf{S}_u$ and semantic consistency with respect to $I_N$.

\subsection{RAGAR}

Our proposed method, RAGAR, is illustrated in Fig.~\ref{fig:overview}. To mitigate the influence of irrelevant historical items, the retrieval module computes semantic similarity scores between the reference item and historical items by correlation unit. The fusion unit then fuses the visual features of the most relevant items, weighted by similarity scores, to construct a retrieval-augmented preference. The generation module employs a LLM to extract and summarise item keywords from the sequence. The LLM and CLIP are then used to derive detailed and global user preferences, respectively, which are subsequently integrated via a balance calibrator. A diffusion-based generator then synthesizes personalized images based on the general preference. Finally, the reflection module evaluates the generated images and updates the model by balancing semantic consistency and personalization through multi-dimensional assessments, thereby ensuring high-quality results.

\begin{figure*}[t]
    \centering
    \includegraphics[width=\textwidth]{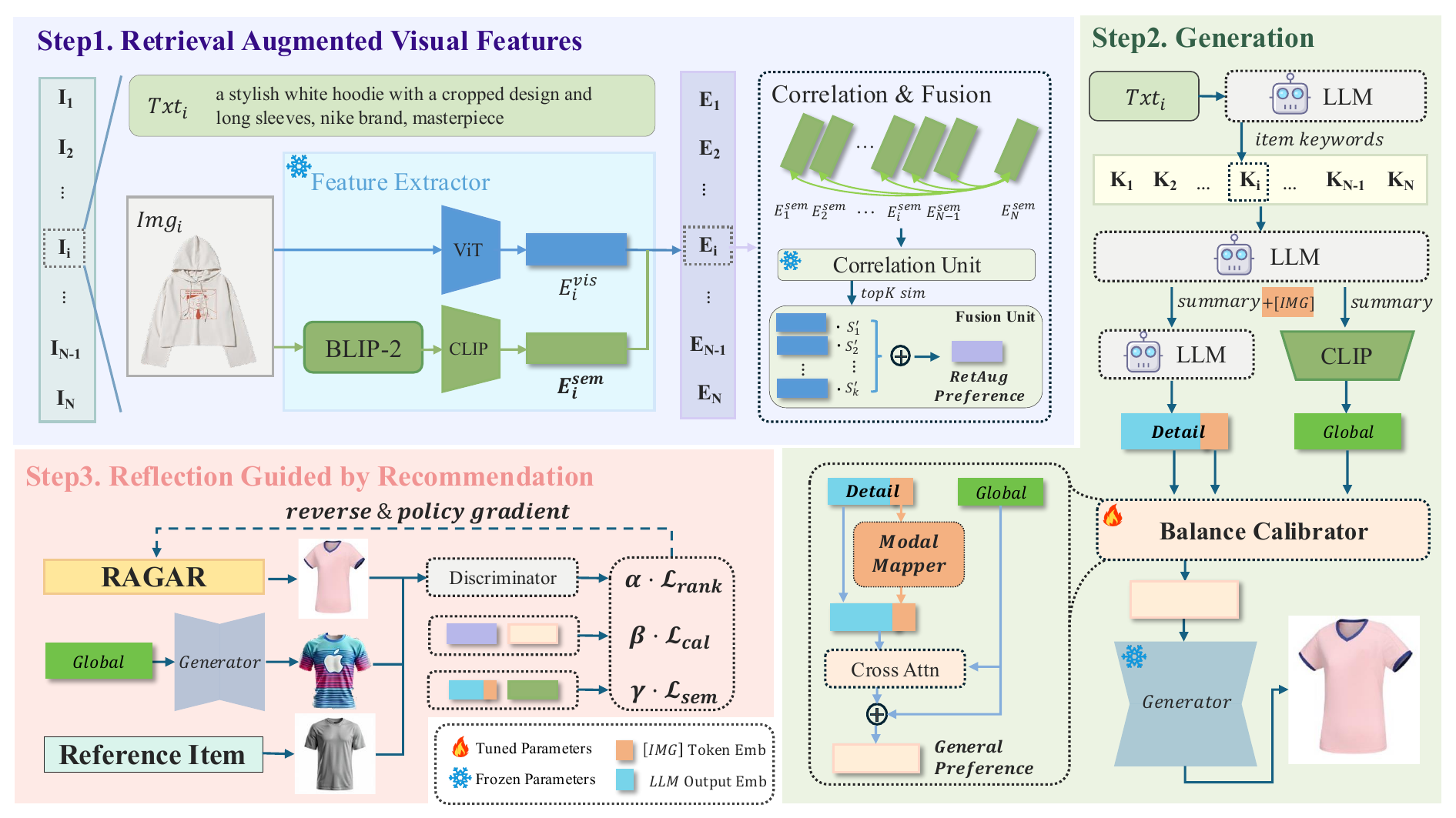}
    % \vspace{-1.5em}
    \caption{\label{fig:overview}\textbf{Overview of the RAGAR framework.} The framework consists of three stages. In the \textbf{Retrieval} stage, semantic and visual features are extracted to identify relevant historical items, whose visual features are fused into the retrieval-augmented preference. The \textbf{Generation} stage uses a LLM, CLIP and the balance calibrator to extract and integrate detailed and global preferences, enabling personalized image generation. In the \textbf{Reflection} stage, the generated images are evaluated through ranking, calibration, and semantic alignment, with policy gradients guiding the model to balance personalization and semantic fidelity.}
    % \vspace{-0.5em}
    \label{fig:overview}
\end{figure*}

\subsection{Retrieval Module}\label{sec:retrieval module}
Given the user’s interaction history $\mathbf{S}_u$ and a reference item $I_N$, we compute the semantic relevance of each historical item. First, we transform items' image into the caption ${cap}_i$ using a pretrained image-to-text model (e.g., BLIP-2~\cite{blip2}).
% \begin{equation}
% cap_i = \textbf{Caption}(img_i), \quad i = 0,1, \cdots, N
% \end{equation}
Each caption is then encoded into a high-dimensional semantic embedding via a text encoder such as CLIP~\cite{clip}:
\begin{equation}
\mathbf{E}^{sem}_i = \mathbf{Enc}^{sem}(cap_i), \quad i = 0,1, \cdots, N
\end{equation}
We compute the semantic similarity between each historical item and the reference item using cosine similarity:
\begin{equation}
s_i = \frac{\sum \mathbf{E}^{sem}_i \mathbf{E}^{sem}_N}{\sqrt{\sum {\mathbf{E}^{sem}_i}^2} \, \sqrt{\sum {\mathbf{E}^{sem}_N}^2}}, \quad i = 0,1, \cdots, N-1
\end{equation}
The top-$k$ most similar items form the retrieval sequence $\mathbf{S}^{ret}_u$, capturing the most relevant preference signals from the user history. To form a retrieval-augmented preference, we extract visual features of the selected items using a visual encoder (e.g., ViT~\cite{vit}):
\begin{equation}
\mathbf{E}^{vis}_i = \mathbf{Enc}^{vis}(img_i), \quad i = 0,1, \cdots, N-1
\end{equation}
These top-k features are then aggregated using the normalized similarity scores $s^{\prime}_i$ as weights. The resulting vector $\mathbf{P}^{ret}$ captures a preference representation that emphasizes semantically aligned visual cues while suppressing irrelevant information.
\begin{equation}
\mathbf{P}^{ret} = \sum_{i=1}^{k} \frac{\exp(s^{\prime}_i)}{\sum_{j=1}^{k} \exp(s^\prime_j)} \cdot \mathbf{E}^{vis}_i
\end{equation}

\subsection{Generation Module} \label{sec:generation_module}
The generation module bridges global preference with detailed preference to produce personalized images. 
To capture user preferences from interactions, we transform interacted items into structured textual descriptions suitable for LLM analysis. Specifically, we summary items with concise keywords to enhance interpretability and reduce extraneous information. We construct a prompt $p_k$ (all prompts are detailed in Appendix) populated with item text to extract concise keywords with an LLM (e.g., Qwen2.5~\cite{qwen}), denoted as $\phi_k$:
\begin{equation}\label{}
    w_i = \phi_k(p_k(cap_i,txt_i)),  i=0,1,\cdots,N-1
\end{equation}
With the keyword list $w_i$ for item $I_i$, we filter out low-frequency keywords and retain the top-$n$ keywords as $w^{\prime}$. Next, we construct a prompt $p_g$ to capture the general user preference. To enhance the expressivity of the LLM, denoted as $\phi_g$, for image generation, we expand its vocabulary by incorporating $L$ additional special tokens ${[\mathbf{IMG}\{i\}]}_{i=1}^L$. The tokens and keywords $w^{\prime}$ are then populated to $p_g$ to extract preferences.
Given the prompt $p_g$, we divide the output into text-related embedding $\mathbf{E}^{txt}$ and $\mathbf{[IMG]}$-related features $\mathbf{E}^{img}\in\mathbb{R}^{L\times d}$:
\begin{equation}\label{}
    [\mathbf{E}^{txt};\mathbf{E}^{img}] = \phi_g (p_g)
\end{equation}
To bridge the gap between textual description and visual representation in LLMs, we adopt the Modal Mapper, a 4-layer encoder-decoder transformer with trainable queries and two linear layers, to align $E^{img}$ with the image space following Gill~\cite{gill}. Then we concatenate the text-related embedding and the image-related embedding and obtain the detailed multi-modal feature.
\begin{equation}\label{}
\mathbf{E}^{d} = [\mathbf{E}^{txt};\phi_m(\mathbf{E}^{img})]
\end{equation}
To maximize keyword utilization and capture the global features, we use the text encoder to generate keyword feature $\mathbf{E}^{g}$. We employ a cross-attention layer and a residual connection to integrate $\mathbf{E}^d$ with $\mathbf{E}^{g}$. In this way, we obtain the general preference feature $P^{gen}$. The generator is then used to generate personalized image.
\begin{equation}\label{}
    \mathbf{P}^{gen} = [\mathrm{softmax} ( \frac{\mathbf{E}^d {\mathbf{E}^{g}}^\top}{{\sqrt{d_{\mathbf{E}^{g}}}}} ) \mathbf{E}^{g}; \mathbf{E}^{g}]
\end{equation}
% \begin{equation}\label{}
%         {v}^{gen} = G_{\psi}(\mathbf{P}^{gen})
% \end{equation}

% TODO(lr)
% \paragraph{Image Generation.}
% With the corrected preference, we utilize the diffusion-based generator to generate images. During training, we sample $r$ random noises $\{\epsilon_t\}_{t=1}^{r} \sim \mathcal{N}(\mu,\sigma^2 I)$ and apply the reparameterization trick to enable gradient backpropagation through the sampling process:
% \begin{equation}
%     \mathbf{z_t} = \mu + \sigma \odot \epsilon_t
% \label{equ:reparameterization}
% \end{equation}
% The generator then produces $r$ images using the policy gradient update strategy discussed in Sec. \ref{sec:reflection module}. Additionally, the keyword feature $\mathbf{E}^{key}$ is used to generate a reference-like image, which serves as input to the ranking model for comparison with the personalized images during the reward computation.

% TODO(lr)
\subsection{Reflection Module}\label{sec:reflection module}
As the diffusion-based generator, propagating gradients backwards from the generated images is challenging \cite{pmg}. To address this issue, we design a multi-assessment reflection module based on policy gradient~\cite{policygradient}, ensuring both user alignment and semantic consistency.

\paragraph{Recommendation Reflection}\label{par:recommendation reflection}
To provide personalized feedback on generated images, we leverage a pre-trained multi-modal ranking model (e.g., MICRO~\cite{micro}) to evaluate images, reducing the reliance on expensive manual labeling. More details about the ranking model (\textbf{RM}) can be found in Appendix.
Given a set of images $ \{v^{ref}, v^{glob}, v^{gen}\}$, where $v^{ref}$ represents the reference image, $v^{glob}$ represents the image generated with the global preference and $v^{gen}$ represents the image generated by RAGAR. By substituting visual features from this set into the sequence of original items, the RM assigns the scores $\rho \in \{\rho^{ref}, \rho^{glob}, \rho^{gen}\}$  and rankings $rk \in \{1,2,3\}$ derived from the score for each item:
\begin{equation}\label{}
    \{\rho, rk\} = \textbf{RM}(S_u, v)
\end{equation}
Intuitively, images that better align with the user’s personalization preference are expected to achieve higher scores and lower ranks.

% To propagate the gradients backwards from images, we designed a way similar to policy gradients.
% The foundation of policy gradient is the Markov Decision Process (MDP), defined as $\langle \mathcal{S},\mathcal{A},\mathcal{R},\mathcal{P},\gamma \rangle$ (state/action spaces, reward, transition probability, discount factor). In RAGAR, we reduce multi-step RL decisions to single-step generation by setting $\mathcal{P}=\gamma=1.0$ while preserving the policy gradient framework. The model parameters serve as states, preference-based image generation as actions, and rank reward as action values. Although states do not change explicitly in this MDP, model parameter updates implicitly induce state transitions. To enable multi-action generation, we sample $r$ Gaussian noise $\epsilon_t,t\in[1, r]$ perturbations on user preferences, treating the mean rank reward of generated images as the final reward. As policy gradient requires gradient ascent, whereas other losses need descent, we reformulate the reward as a penalty term to be minimized by taking the negative of the formula.

To enable gradient propagation from image-level feedback, we adopt a policy gradient-inspired approach. The underlying formulation is based on a Markov Decision Process (MDP), defined as $\langle \mathcal{S}, \mathcal{A}, \mathcal{R}, \mathcal{P}, \gamma \rangle$ (state/action spaces, reward, transition probability, discount factor). In RAGAR, we simplify the multi-step decision process into a single-step generation by setting $\mathcal{P} = \gamma = 1.0$, while retaining the core policy gradient framework. Model parameters are treated as states, image generation conditioned on user preference as actions, and rank-based rewards as action values. Although the states do not transition explicitly, parameter updates implicitly induce state changes. To support multi-action exploration, we sample $r$ perturbations $\epsilon_t$, $t \in [1, r]$, from a Gaussian distribution over user preferences and compute the final reward as the mean rank score of the generated images. Since policy gradients require maximizing reward while other losses are minimized, we incorporate the reward as a penalty term by negating it.

Specifically, we employ a reward function to guide the model toward generating images that reflect personalization preferences. We then accumulate rewards for each noise $\epsilon_t$ and calculate the $\mathcal{L}_{rank}$.
\begin{equation} \label{equ:rank reward}
        \mathcal{R}_t = \sum_{c\in \{ref,glob\}}\max\left(0,\rho^c-\rho^{gen}_t\right) + \delta
\end{equation}
\begin{equation}\label{}
        \mathcal{L}_{rank} =-\frac{1}{r}\sum_{t=1}^r \log_{\pi_\theta}(\mathbf{P}^{gen} + \epsilon_t)\cdot \mathcal{R}_t
\end{equation}
To encourage the model to generate higher-quality images we add a margin $\delta$. And $\log_{\pi_\theta}$ computes the probability density of the generated embedding under the Gaussian distribution.

% Moreover, the foundation of Policy Gradient is the Markov Decision Process, defined as $\langle \mathcal{S},\mathcal{A},\mathcal{R},\mathcal{P},\gamma \rangle$ (state/action spaces, reward, transition probability, discount factor). In RAGAR, we reduce multi-step RL decisions to single-step generation by setting $\mathcal{P}=\gamma=1.0$ while preserving the Policy Gradient framework. And model parameters serve as states, preference-based image generation as actions, and rank reward as action values. Although states do not change explicitly in this MDP, model parameter updates implicitly induce state transitions. To enable multi-action generation, we sample $N$ Gaussian noise perturbations on user preferences, treating the mean rank reward of generated images as the final reward. As Policy Gradient requires gradient ascent, whereas other losses need descent, we reformulate the reward as a penalty term to be minimized by taking the negative of the formula.
% By incorporating personalization reflection, the model learns to generate images that better reflect user preferences and exhibit features relevant to the interaction history.

\paragraph{Calibrator Reflection}
As mentioned in Sec. \ref{sec:retrieval module}, the retrieval-augmented preference focuses primarily on item features associated with the reference item. To narrow the gap between the general preference feature $\mathbf{P}^{gen}$ and the retrieval-augmented preference feature $\mathbf{P}^{ret}$, we minimize the calibrator loss between them:
\begin{equation}\label{}
\mathcal{L}_{\mathrm{cal}} = \Vert \mathbf{P}^{gen} - \mathbf{P}^{ret} \Vert_2^2
\end{equation}
This adjustment ensures that $P^{gen}$ reflects both the general and retrieval-augmented preferences, and guide the model to learn from retrieval.

\paragraph{Semantic Reflection}
To enhance the semantic consistency, we minimize the distance between the modal mapper's output in the generation module, $\mathbf{E}^{d}$, and the semantic features of the reference image, $\mathbf{E}^{sem}_N$, using the semantic loss. Combined with the semantic loss, generated images align more semantically with the reference.
\begin{equation}\label{}
\mathcal{L}_{\mathrm{sem}} = \Vert \mathbf{E}^{d}  - \mathbf{E}^{sem}_N \Vert_2^2
\end{equation}

\paragraph{Reflection Loss}
We update the Balance Calibrator in the generation module. The overall loss is defined with three adjustable weighting hyper-parameters $\alpha$, $\beta$ and $\gamma$:
\begin{equation}\label{equ:joint reflection}
\mathcal{L} =\alpha \cdot \mathcal{L}_{rank} + \beta \cdot \mathcal{L}_{\mathrm{cal}}+\gamma \cdot \mathcal{L}_{\mathrm{sem}}
\end{equation}
% By optimizing this joint loss, the RAGAR framework effectively balances personalization with semantic consistency, enabling the generation of high-quality, personalized images.

\section{Experiment} \label{sec:experiment}
We evaluate RAGAR on three different scenarios: commodity, movie poster, and sticker. Our experiments aim to answer the following research questions:
\begin{itemize}
\setlength{\itemsep}{0.5em}    % 控制item之间的间距
\setlength{\parskip}{0.3em}    % 控制段落之间的间距
\item \textbf{RQ 1:} How does RAGAR perform compared to other generative methods in quantitative evaluation metrics? 
\item  \textbf{RQ 2:} Does RAGAR outperform baseline methods in human evaluation?
\item \textbf{RQ 3:} What is the impact of each module of RAGAR on performance? Specifically, how do the retrieval module and the reflection module contribute to the generation?
\item \textbf{RQ 4:} Can personalized images generated by RAGAR improve the performance in recommendation systems?
\end{itemize}

\begin{table*}[!htb]
    \centering
    \small
    \renewcommand{\arraystretch}{1.2}
\renewcommand{\tabcolsep}{3pt}

\begin{tabular}{c|c|ccccc|cccc} 
\toprule[1.1pt]
\multirow{2}{*}{\textbf{Datasets}}  & \multirow{2}{*}{\textbf{Methods}} & \multicolumn{5}{|c|}{\textbf{Personalization}}    & \multicolumn{4}{|c}{\textbf{Semantic Alignment}}  \\
&& $\Delta{R}\uparrow$  & CPS$\uparrow$ & CPIS$\uparrow$    & LPIPS$\downarrow$ & SSIM$\uparrow$    & CS$\uparrow$  & CIS$\uparrow$     & LPIPS$\downarrow$ & SSIM$\uparrow$ \\
\midrule
\multirow{7}{*}{POG}
& GLIDE & -54.02    & 13.90    & 59.01    & \underline{49.01} & 17.24 & 17.52 & 59.17 & \underline{58.35}   & \textbf{27.14}   \\
& SD-v1.5   & -25.88    & 15.02 & 62.97 & 55.46 & 12.53 & \underline{24.18} & 63.72  & 63.50 & 12.39   \\
& TI     & -25.64  & \underline{15.64} & \underline{63.79} & 53.68 & 15.67 & 24.14 & 67.83 & 62.40 & 15.23  \\
& LaVIT & -51.09  & 15.40 & 63.46 & \textbf{48.77} & \textbf{22.65} & 24.07 & \underline{72.04} & \textbf{56.42} & \underline{24.79}          \\
& PMG      & \underline{-22.96}  & 14.82 & 56.74 & 55.76 & 5.19 & 18.48 & 57.00 & 63.89 & 5.02          \\
& \textbf{RAGAR}     & \textbf{-19.77}  & \textbf{15.79} & \textbf{66.88} & 55.93 & \underline{17.99} & \textbf{24.28} & \textbf{74.55} & 59.29 & 19.08          \\
               
        \midrule
        \multirow{7}{*}{ML-latest}     
                & GLIDE   & -3.20  & 13.27 & 29.52 & 60.34 & \underline{17.99} & 18.21 & 42.31 & 61.88 & \textbf{26.22}          \\
                
                & SD-v1.5   & -1.97  & 12.85 & 31.07 & 59.74 & 14.02 & 18.38 & 53.59 & 58.58  & 14.51   \\
           
                & TI     &  -2.58  & 14.42 & 32.85 & 59.33 & 14.29 & \textbf{25.25} & \underline{54.17} & 59.35 & 14.79         \\
                     
                & LaVIT & -1.54  & 14.80 & 34.78 & 57.01 & \textbf{18.16} & \underline{21.04} & \textbf{57.56} & 56.35 & \underline{19.64}          \\

                & PMG    & \underline{-0.11}  & \underline{15.07} & \underline{41.80} & \textbf{53.83} & 7.34 & 14.30 & 43.85 & \textbf{53.51} & 7.29          \\
            
                & \textbf{RAGAR}    & \textbf{0.01}  & \textbf{16.20} & \textbf{43.45}   & \underline{56.56} & 15.14 & 19.04 & 53.24 & \underline{56.30} & 15.75          \\
        \midrule
        \multirow{7}{*}{Sticker}
                & GLIDE   & -0.86  & 12.27 & 49.52 & 59.78 & 18.00 & 16.47 & 48.97 & 67.88 & \textbf{28.23} \\
               & SD-v1.5     & -0.93  & 12.48 & 51.08 & 59.45 & 17.38 & 18.46 & 49.61 & 68.41 & 16.43          \\
                & TI   & -0.83  & 13.24 & 50.85 & 59.48 & 16.76 & 17.26 & 50.14 & 67.17 & 15.72   \\

                & LaVIT & -0.75  & 12.64 & \underline{53.05} & \underline{57.82} & \underline{24.41} & \textbf{21.12} & \underline{58.99} & \underline{62.27} & 24.43          \\
           
                & PMG      & \underline{-0.73}  & \underline{13.31} & 50.97 & 59.93 & 4.94 & \underline{18.72} & 50.92 & 68.81  & 4.73          \\
            
                & \textbf{RAGAR}     & \textbf{0.99}  & \textbf{14.97} & \textbf{55.25} & \textbf{56.71} & \textbf{25.61} & 17.69 & \textbf{64.41} & \textbf{61.38} & \underline{27.48}       \\
        \bottomrule
    \end{tabular}
    \caption{\textbf{Quantitative performance comparison on three datasets in terms of personalization and  semantic alignment.} The best performance is \textbf{bold} while the second-best is \underline{underlined}.}
    \label{tab:performence comparison}
    % \vspace{-2.0em}
\end{table*}

\subsection{Experimental Settings}

% \begin{wraptable}{r}{0.4\textwidth}
%     \centering
%     \small
%     \renewcommand{\arraystretch}{1.2}
%     \renewcommand{\tabcolsep}{3pt}
%     \caption{\textbf{Characteristics of experimental datasets.}}
%     \label{tab:dataset details}
%     \vspace{0.5em}
% \begin{tabular}{cccc}
%     \toprule
%     \textbf{Features}&\textbf{POG} & \textbf{ML-latest} & \textbf{SER30K} \\
%     \midrule
%     \textbf{\#Users}   & 1000 & 4689 &  2230 \\      
%     \textbf{\#Items}  & 19242 & 9742 & 30739 \\      
%     \bottomrule
% \end{tabular}
% \end{wraptable}

\begin{table}[h]
    \centering
    \small
    \renewcommand{\arraystretch}{1.2}
    \renewcommand{\tabcolsep}{3pt}
    % \vspace{0.5em}
\begin{tabular}{cccc}
    \toprule
    \textbf{Features}&\textbf{POG} & \textbf{ML-latest} & \textbf{SER30K} \\
    \midrule
    \textbf{\#Users}   & 1000 & 4689 &  2230 \\      
    \textbf{\#Items}  & 19242 & 9742 & 30739 \\      
    \bottomrule
\end{tabular}
    \caption{\textbf{Characteristics of experimental datasets.}}
    \label{tab:dataset details}
\end{table}

\paragraph{Datasets.} \label{sec:secnarios_dataset}
We utilize three real-world datasets to generate personalized images across different scenarios: POG, ML-latest ,SER30K. POG is a multi-modal dataset of fashion clothing with user interaction history. Due to the large size, we randomly select a subset of users and interactions. ML-latest is a benchmark movie dataset with user ratings. We collect corresponding posters from IMDB and split user historical sequences to ensure each contains 20 interactions. SER30K is a large-scale sticker dataset where each sticker is categorized by theme and annotated with an associated emotion label. We merge two random themes into user historical sequences and randomly select a reference sticker from other themes. Dataset statistics are summarized in Tab.~\ref{tab:dataset details}. 

% \noindent % 取消段落首行缩进
% \begin{minipage}[t]{0.48\textwidth} % 左侧文本区域，占页面宽度的 48%
% \end{minipage}
% \hfill % 添加一些水平间距
% \begin{minipage}[t]{0.48\textwidth} % 右侧表格区域，占页面宽度的 48%
% % \input{tables/dataset_details}
%     \centering
%     \small
%     \renewcommand{\arraystretch}{1.2}
%     \renewcommand{\tabcolsep}{3pt}
%     % \caption{The characteristics of experimental datasets.}
%     \label{tab:dataset details}
% \begin{tabular}{cccc}
%     \toprule
%     \textbf{Features}&\textbf{POG} & \textbf{ML-latest} & \textbf{SER30K} \\
%     \midrule
%     \textbf{\#Users}   & 1000 & 4689 &  2230 \\      
%     \textbf{\#Items}  & 19242 & 9742 & 30739 \\      
%     \bottomrule
% \end{tabular}
% \end{minipage}

% For POG, due to the large dataset size, we selected a subset of users and their interaction histories to generate personalized product images. For ML-latest, we collect the corresponding movie posters from IMDB\footnote{https://www.imdb.com.} and split the user sequence to create more interaction histories, ensuring each sequence contains approximately 20 interactive items. For SER30K, we merged two random themes into a user interaction history sequence, randomly selected a sticker from other themes as the reference item, and placed it at the end of the sequence. Dataset statistics are summarized in Table \ref{tab:dataset details}. 

\paragraph{Comparison Methods.}
We compare RAGAR with five generative baselines, including three DM-based models Glide~\cite{glide}, SD~\cite{ldm}, and TI~\cite{ti}, and two LLM-based models LaVIT~\cite{lavit} and PMG~\cite{pmg}.

\paragraph{Evaluation Metrics.}
We compare RAGAR with baselines through the following quantitative metrics: To evaluate \textbf{semantic alignment}, we calculate CLIP Score (CS), CLIP Image Score (CIS), LPIPS \cite{lpips} and SSIM \cite{ssim} for reference images.
To evaluate \textbf{personalization}, we calculate the CLIP Personalization Score (CPS) and the CLIP Personalization Image Score (CPIS), which measure the similarity between the generated images and the text description or images representing user preferences, respectively. In addition, we calculate the LPIPS \cite{lpips} and SSIM \cite{ssim} to quantify the perceptual similarity. Furthermore, we compute the rank change $\mathbf{\Delta R}$ between the original and generated images. This metric emphasizes improvements for higher-ranked items by diminishing the weight for lower-ranked ones. A larger $\mathbf{\Delta R}$ indicates greater personalization in the image.
\begin{equation}\label{}
\mathbf{\Delta R} = \frac{rk_{ori} - rk_{gen}}{1 + rk_{ori}}
\end{equation}

\paragraph{Parameter Settings.} \label{par:param_settints}
To make a fair comparison, all baselines are tuned with a fixed learning rate of $1e^{-5}$ and Stable Diffusion 1.5 is used as the image generator.
For RAGAR, we set the learning rate at $1e^{-5}$. The number of retrieval items is fixed to 5. The number of noise is set at 3. The hyperparameters for joint reflection are $\alpha=0.2, \beta=0.5, \gamma=0.3$. More training details can be found in the Appendix.
All experiments are conducted on a single NVIDIA-A100 GPU.

\subsection{Performance Comparison (RQ1)}
Tab.~\ref{tab:performence comparison} presents the comparison between RAGAR and baseline methods, from which several key observations can be made. First, traditional DM-based methods show poor performance on personalization metrics. Glide and SD rank lowest across all five personalization metrics on every dataset, largely because both rely on CLIP for image generation, which struggles to capture deeper semantic associations. TI, which augments SD with stylized word embeddings to model user preferences, significantly improves personalization (e.g., $\Delta R$: -25.64 vs. -54.02, -2.58 vs. -3.20, -0.83 vs. 0.86). In terms of semantic alignment, all three methods perform well, with Glide achieving the highest SSIM, indicating a tendency to reconstruct reference images.

Second, LLM-based methods perform better overall, owing to their stronger textual understanding. LaVIT, a multi-modal transformer, better aligns visual and textual features but still underperforms in personalization (e.g., $\Delta R$: -51.09 on POG). PMG, which extracts user preferences via LLMs, achieves better personalization than DM-based baselines but remains constrained by its dependence on reference image consistency.

Finally, RAGAR achieves state-of-the-art performance across all datasets, benefiting from sequence retrieval and a ranking-based training strategy. The gain is especially notable on the ML dataset, likely due to its larger volume of front-end user data. Compared to PMG, RAGAR further improves consistency through semantic retrieval and the use of evaluation metrics more aligned with human judgment, leading to superior personalization.

\begin{table}[!htb]
\centering
\small
\renewcommand{\arraystretch}{1.2}
\renewcommand{\tabcolsep}{3pt}
    % \begin{minipage}{0.48\textwidth}
    \centering
    \small
    \label{tab:user study}
    % \vspace{0.5em}
    \begin{tabular}{ccccccc}
    \toprule
    \multirow{2}{*}{Methods} & \multicolumn{2}{c}{POG}  & \multicolumn{2}{c}{ML-latest} & \multicolumn{2}{c}{Sticker}   \\
    \cmidrule(l){2-7}
    & Per.  & Sen.  & Per.  & Sen.  & Per.  & Sen.  \\ 
    \midrule
    ORI & 2.26  & /     & 2.16  & /     & 2.18  & /     \\
    PMG & 2.08  & 1.62  & 2.10  & 1.94  & 2.24  & 1.88  \\
    \textbf{RAGAR}  & \textbf{1.66} & \textbf{1.46} & \textbf{1.74} & \textbf{1.66} & \textbf{1.58} & \textbf{1.12} \\ 
    \bottomrule
    \end{tabular}
    \caption{\textbf{Human evaluation of RAGAR, PMG and origin images.} RAGAR consistently outperforms both ORI and PMG in all metrics, achieving up to 29.46\% improvement in personalization and 40.43\% in semantic alignment.}
    \label{tab:user study}
\end{table}
% \vspace{-1.5em}

\subsection{Human Evaluation (RQ2)}\label{subsec:human_eval}
To evaluate RAGAR's effectiveness in personalization and semantic alignment, we conducted a human study comparing it with PMG and original images. Fifty participants completed two ranking tasks (50 cases per dataset): one on personalization—ranking generated images by their likelihood of being clicked based on 5 historical items—and one on semantics—ranking images by how well they convey the meaning of a reference image. As shown in Tab.~\ref{tab:user study}, RAGAR consistently outperforms baselines on both tasks, demonstrating its superior ability to capture user preferences while preserving semantic content.
% To assess the personalization and semantics of RAGAR in real scenarios, we conduct a human evaluation to compare it with PMG and original images. We invited 50 volunteers and set up two types of sorting tasks, with 50 cases for each dataset, covering personalization and semantics. For personalization, given 5 historical images along with the generated images, participants are asked to sort images in descending order of likelihood that the user would click them, based on historical click data. For semantics, given the reference image and generated images, participants are asked to sort them in descending order of how well they convey the same meaning as the reference image. As shown in Tab. \ref{tab:user study}, RAGAR outperforms the baselines on both metrics. This indicates RAGAR better reflects user preference while preserving the semantics.

\subsection{Ablation Study (RQ3)}
We study the effectiveness of key components of RAGAR in the POG dataset, including the retrieval and reflection module. Experiments on more datasets can be found in the Appendix.
\textbf{Effect of retrieval module.}
We assess the impact of the retrieval module on improving semantic consistency and capturing user preferences by excluding it during training. Fig. \ref{exp_abl}(a) shows that: 1) noisy items in the historical sequence diminish the model's ability to capture user preferences and interfere with semantics; 2) the retrieval module selects items relevant to user preferences, thereby improving the performance.
Furthermore, we investigate the effect of varying the retrieval number $k$ during training. The results reveal that $k=5$ yields the best results when the historical sequence length is 20. At $k=0$ (no retrieval), performance is poor, while increasing $k$ initially improves performance but declines beyond $k=5$ due to reintroduced noise and redundancy. 
\textbf{Effect of rank rewards.} 
We investigate the role of rank rewards by excluding it while retaining the other loss functions during training, denoted as w/o. The results in Fig. \ref{exp_abl}(b) demonstrate that the model's performance decreases without rank rewards. Adding the amount $r$ of sample noise $\epsilon$ boosts the model's performance. Specifically, $r=5$ strikes the optimal balance between personalization and semantic alignment.

\begin{figure}[t]
    \centering
    \includegraphics[width=\linewidth,scale=1.00]{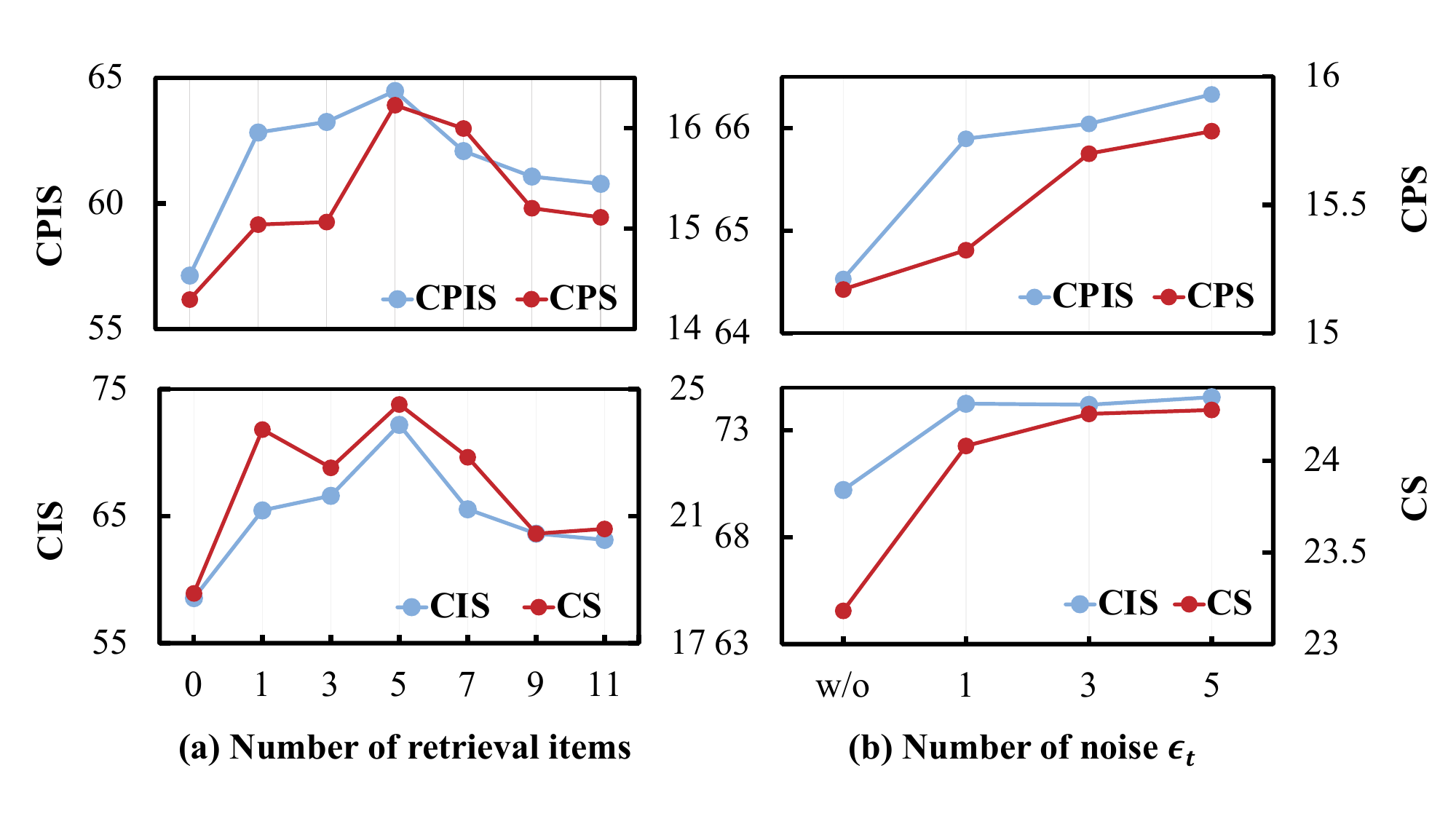} 
    \vspace{-1.5em}
    \caption{\label{exp_abl}  \textbf{The impact of the retrieval module and the rank reward mechanism.} Performance improves as the number of retrieval items and noise samples increases, peaking at moderate values (around 5 for retrieval and noise), after which excessive inputs introduce noise and degrade results.}
    \vspace{-1.5em}
    \label{exp_abl}
\end{figure}

% \begin{wraptable}{r}{0.48\textwidth}
%     \centering
%     \small
%     \renewcommand{\arraystretch}{1.2}
% \renewcommand{\tabcolsep}{3pt}
% % \caption{Human evaluation comparison on both personalization and semantic alignment.}
% \caption{\textbf{Human evaluation of RAGAR, PMG and the origin image.} RAGAR consistently outperforms both ORI and PMG across all metrics, achieving up to 29.46\% improvement in personalization and 40.43\% in semantic alignment over PMG.}
% \label{tab:user study}
% \vspace{0.5em}
% \begin{tabular}{ccccccc}
% \toprule
% \multirow{2}{*}{Methods} & \multicolumn{2}{c}{POG}  & \multicolumn{2}{c}{ML-latest} & \multicolumn{2}{c}{Sticker}   \\
% \cmidrule(l){2-7}
% & Per.  & Sen.  & Per.  & Sen.  & Per.  & Sen.  \\ 
% \midrule
% ORI & 2.26  & /     & 2.16  & /     & 2.18  & /     \\
% PMG & 2.08  & 1.62  & 2.10  & 1.94  & 2.24  & 1.88  \\
% \textbf{RAGAR}  & \textbf{1.66} & \textbf{1.46} & \textbf{1.74} & \textbf{1.66} & \textbf{1.58} & \textbf{1.12} \\ 
% \bottomrule
% \end{tabular}
% \end{wraptable}

\begin{table}[!htb]
    \centering
    \small
    \label{tab:auxiliary generation}
    % \vspace{0.5em}
    \begin{tabular}{ccccc}
    \toprule
    \multirow{2}{*}{Methods} & \multicolumn{2}{c}{POG}           & \multicolumn{2}{c}{ML-latest}     \\ \cmidrule(l){2-5} 
                             & R@10            & N@10            & R@10            & N@10            \\ \midrule
    ORI                      & 0.1765          & 0.1544          & {0.3229}    & 0.2653          \\
    PMG                      & {0.1804}    & {0.1550}    & 0.3213          & {0.2682}    \\
    \textbf{RAGAR}           & \textbf{0.1927} & \textbf{0.1668} & \textbf{0.3299} & \textbf{0.2882} \\ \bottomrule
    \end{tabular}
    \caption{\textbf{Application of personalized generation in recommendation domain}. Compared to PMG and original images(ORI), RAGAR leads to an average improvement of approximately 7\% in recommendation metrics. }
    \label{tab:auxiliary generation}
    \vspace{-0.5em}
\end{table}
\vspace{-0.5em}

\begin{figure*}[!h]
    \centering
    \includegraphics[width=1.0\textwidth]{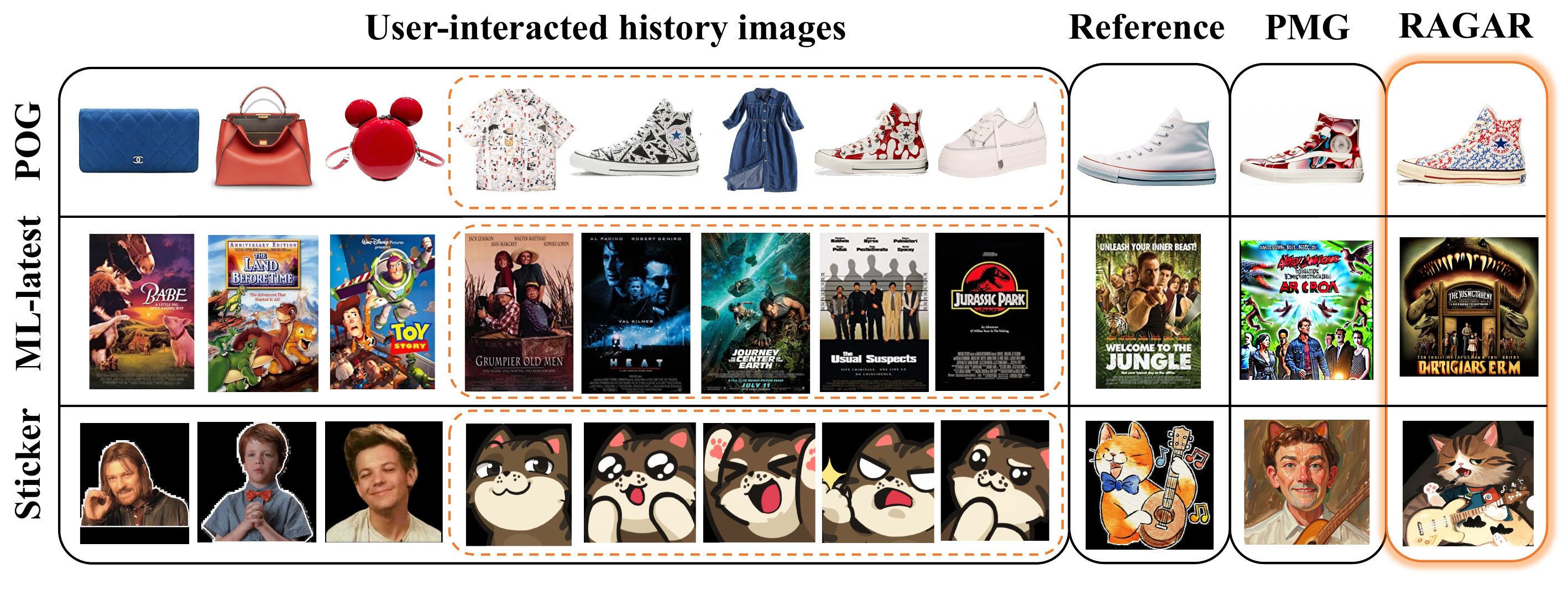}
    % \vspace{-1.5em}
    \caption{\label{fig:case study}\textbf{Qualitative comparison of different methods.} Three representative cases highlight how RAGAR incorporates retrieved features into the generation. Left: user interaction history (dashed boxes indicate retrieved items). Right: images generated by RAGAR and PMG.}
    \vspace{-1.5em}
    \label{fig:case study}
\end{figure*}

\subsection{Auxiliary Generation (RQ4)}
Generating items beyond the original data distribution can reveal user interests not captured by existing data \cite{surver4gene-rec}. RAGAR-generated personalized images not only offer high visual quality but also improve recommendation performance. We validate this using the multi-modal recommendation model MICRO \cite{micro} on two real-world datasets: POG and ML-latest.
In our experiments, we replace reference images with those generated by RAGAR, PMG, and original images (ORI) during rank model training. As shown in Tab.~\ref{tab:auxiliary generation}, both PMG and RAGAR outperform ORI across all metrics, highlighting the advantage of generative approaches. RAGAR achieves the best results, improving Recall@10 and NDCG@10 by 6.8\% and 7.6\% on POG, and by 2.67\% and 7.4\% on ML-latest, demonstrating its superior ability to model user preferences.

\subsection{Case Study}
We showcase personalized image generation results from RAGAR and PMG on the POG, ML-latest, and Sticker datasets (Fig.~\ref{fig:case study}). Each example includes eight historical items, a reference image, and generated outputs, with orange dashed boxes marking items retrieved by RAGAR.
In the clothing case, RAGAR produces red-and-blue shoes that match user preferences while filtering out irrelevant styles like patent leather, outperforming PMG. In the movie poster case, RAGAR reflects the user’s preference for multiple characters and avoids unrelated anime elements, which PMG fails to do. In the sticker case, RAGAR maintains semantic consistency (e.g., cat and guitar) and style (e.g., pink ears, brown fur), while excluding unwanted realistic styles. More examples are in the Appendix.

\section{Related Works}
\paragraph{Diffusion models (DMs)} \cite{ldm} have enabled high-quality image generation \cite{glide,sgdm,pgr,jedi,wang2024qihoo,zhang2025u,he2025plangen,lu2025uni,bi2024using,zhang2024towards} and video generation \cite{bi2025customttt,xu2025dropoutgs,wang2025wisa,cao2025relactrl,feng2024fancyvideo,shao2025eventvad}, but personalization remains underexplored. Early methods \cite{ti,dreambooth} rely on text prompts, while others \cite{difashion,cg4ctr,adabooster} combine user interactions and prompts to generate personalized content. Some approaches connect DMs with large language models (LLMs) \cite{gpt4,llama,qwen}, such as \cite{gill}, and \cite{pmg}. However, \cite{pmg} depends solely on consistency loss, often overfitting the reference image and ignoring user preferences. Multi-modal LLMs \cite{llava,gemini,lavit} generate personalized images from text but lack fine-grained preference modeling.

\paragraph{Recommendation systems} aim to deliver content aligned with user preferences, using retrieval \cite{rag,MORE,wang2025learning} and ranking methods \cite{rsinllm,vip5,micro}. Multi-modal approaches like \cite{MORE} fuse text and image features for better reasoning, while \cite{vip5} adds visual understanding. \cite{micro} further models cross-modal user-item relationships. More detailed related work provide in the Appendix.

% \section{Limitations} \label{sec:limitations} % 写更多内容
% First, our method relies on a pre-trained recommendation system during training, which may introduce dependency on its quality and coverage. Second, to maintain plug-and-play compatibility, we do not alter the generator architecture, which may occasionally lead to minor inconsistencies between the preference representation and the generated output. Finally, in cases where the user’s interaction history is weakly related to the reference image, the retrieval module may still return top-$k$ items with limited relevance, potentially reducing the effectiveness of personalization.

\section{Conclusion} \label{sec:conclusion}
% We propose and validate two key assumptions in Sec.~\ref{sec:intro} and introduce RAGAR, a framework that enhances personalization through retrieval-augmented preference and recommendation while preserving semantic consistency. A Balance Calibrator combines global and detailed preferences to guide generation, and recommendation signals with policy gradients address image-level gradient propagation. Future work will inject user preferences directly into the diffusion process for better alignment with user intent.
In the paper, we propose and validate two key assumptions in Sec.~\ref{sec:intro}. Based on these insights, we introduce RAGAR, a novel framework that enhances personalization via retrieval-augmented preference and recommendation while preserving semantic consistency. We propose the Balance Calibrator to combine global and detailed preferences for guiding personalized generation. To address gradient propagation at the image level, we integrate recommendation signals and policy gradients into training. In future work, we aim to inject user preferences directly into the diffusion process to better align generation with user intent.

\bibliography{xxxx2026}
% \newpage
% \clearpage
% \input{./checklist}

\end{document}